# A new quasi-one-dimensional superconductor parent compound NaMn$_6$Bi$_5$ with lower antiferromagnetic transition temperatures

Ying Zhou[1, 2, #], Long Chen[1, 2, #], Gang Wang[*, 1, 2, 4], Yuxin Wang[1, 2], Zhichuan Wang[1, 2], Congcong Chai[1, 2], Zhongnan Guo[3], Jiangping Hu[*, 1, 2, 4], Xiaolong Chen[*, 1, 2, 4]

[1] Beijing National Laboratory for Condensed Matter Physics, Institute of Physics, Chinese Academy of Sciences, Beijing 100190, China

[2] University of Chinese Academy of Sciences, Beijing 100049, China

[3] Department of Chemistry, School of Chemistry and Biological Engineering, University of Science and Technology Beijing, Beijing 100083, China

[4] Songshan Lake Materials Laboratory, Dongguan, Guangdong 523808, China

[#] These authors contributed equally to this work.

[*]Corresponding author: gangwang@iphy.ac.cn; jphu@iphy.ac.cn; xlchen@iphy.ac.cn;

**Abstract:** Mn-based superconductor is rare and recently reported in quasi-one-dimensional KMn$_6$Bi$_5$ with [Mn$_6$Bi$_5$]$^-$ columns under high pressure. Here we report the synthesis, magnetic properties, electrical resistivity, and specific heat capacity of the newly-discovered quasi-one-dimensional NaMn$_6$Bi$_5$ single crystal. Compared with other AMn$_6$Bi$_5$ (A = K, Rb, and Cs), NaMn$_6$Bi$_5$ has larger intra-column Bi-Bi bond length, which may result in the two decoupled antiferromagnetic transitions at 47.3 K and 52.3 K. The relatively lower antiferromagnetic transition temperatures make NaMn$_6$Bi$_5$ a more suitable platform to explore Mn-based superconductors. Anisotropic resistivity and non-Fermi liquid behavior at low temperature are observed. Heat capacity measurement reveals that NaMn$_6$Bi$_5$ has similar Debye temperature with those of AMn$_6$Bi$_5$ (A = K and Rb), whereas the Sommerfeld coefficient is unusually large. Using first-principles calculations, the quite different density of states and an unusual enhancement near the Fermi level are observed for NaMn$_6$Bi$_5$, when compared with those of other AMn$_6$Bi$_5$ (A = K, Rb, and Cs) compounds.

# 1 Introduction

Finding unconventional superconductors violating the Bardeen-Cooper-Schrieffer (BCS) theory is a promising routine to reach room-temperature superconductivity under ambient pressure and brings out fruitful results in the past decades, like cuprates [1, 2], iron pnictides [3, 4] and heavy-fermion superconductors [5, 6]. Apart from the cuprates and iron pnictides, other $3d$-transition-metal-based compounds are also of great interest in exploring superconductors for their varying coordination environment of $3d$ transition metal [7, 8] and unique magnetism [9, 10]. By suppressing the magnetism of $3d$-transition-metal-based compounds, unconventional superconductivity is expected to be observed in the vicinity of antiferromagnetic quantum critical point (QCP). However, up to now, only a few $3d$-transition-metal-based superconductors, such as Cr-based [11-14] or Mn-based [15] superconductors, are reported.

Mn-based superconductor was first reported in three-dimensional MnP with its helical magnetic order suppressed under pressure [15], which suggests that superconductivity could emerge in Mn-based system by suppressing the helical magnetic order. In our former work, an interesting helical magnetic order was predicted in $RbMn_6Bi_5$ [16], which is a quasi-one-dimensional ternary Mn-based compounds with unique $[Mn_6Bi_5]^-$ columns similar to $KMn_6Bi_5$ [17]. Following $RbMn_6Bi_5$, we then substituted Rb by smaller alkali metal Na and synthesized $NaMn_6Bi_5$ having lower antiferromagnetic transition temperatures. Very recently, J. G. Cheng, et al. has reported a pressure-induced superconductivity with transition temperature up to 9.3 K in $KMn_6Bi_5$ by suppressing the antiferromagnetic order [18], which may open a new avenue for finding more Mn-based superconductors. Considering the even lower antiferromagnetic transition temperatures of $NaMn_6Bi_5$ than that of $KMn_6Bi_5$, we believe that $NaMn_6Bi_5$ is a more suitable platform to explore superconductivity and other possible exotic properties. Here, the single crystal growth, magnetism, and transport properties of $NaMn_6Bi_5$ are reported. For the smaller radius of Na, $NaMn_6Bi_5$ shares a same quasi-one-dimensional structure motif as that of $KMn_6Bi_5$ and $RbMn_6Bi_5$, but having an unusual enhancement of intra-column Bi-Bi bonds. The suppression of antiferromagnetic transition is confirmed by magnetic susceptibility, specific heat capacity, and resistivity measurements. Meanwhile, anisotropic resistivity, non-Fermi liquid behavior, and largely enhanced Sommerfeld coefficient are observed. First-principles calculations reveal a quite different density of states (DOS) near the Fermi level for $NaMn_6Bi_5$ compared with those of other $AMn_6Bi_5$ (A = K, Rb, and Cs). These results demonstrate that $NaMn_6Bi_5$ is a more promising candidate to explore Mn-based superconductor.

## 2 Experimental Methods

**Single Crystal Growth.** $NaMn_6Bi_5$ single crystals were grown by a high-temperature solution method modified from that reported in our former work [16]. Na chunk (99.75%, Alfa Aesar), Mn powder (99.95%, Alfa Aesar), and Bi granules (99.999%, Sinopharm) were mixed using a molar ratio of Na : Mn : Bi = 0.5 : 4 : 8 in a fritted alumina crucible set (Canfield crucible set) [19] and sealed in a fused-silica ampoule at vacuum. The ampoule was heated to 1073 K over 15 h, held at the temperature for 24 h, and then slowly cooled down to 673 K at a rate of 2 K/h. At 673 K, the single crystals with size up to 5 mm × 0.5 mm × 0.5 mm were separated from the remaining liquids by centrifuging the ampoule. The obtained single crystals are shiny-silver needles and air-sensitive, so all manipulations and specimen preparation for structure characterization and property measurements were handled in an argon-filled glovebox. In addition, $CsMn_6Bi_5$ single crystals were also grown by using the similar method.

**Structure Characterization and Composition Analysis.** X-ray diffraction data were obtained using a PANalytical X'Pert PRO diffractometer (Cu $K_\alpha$ radiation, λ = 1.54059 Å) operated at 40 kV voltage and 40 mA current with a graphite monochromator in a reflection mode (2θ = 5°–100°, step size = 0.017°). Indexing and Rietveld refinement were performed using the DICVOL91 and FULLPROF programs [20]. Single crystal X-ray diffraction (SCXRD) data were collected using a Bruker D8 VENTURE with Mo $K_\alpha$ radiation (λ = 0.71073Å) at 300 K. The morphology and analyses of elements were characterized using a scanning electron microscope (SEM, Hitachi S-4800) equipped with an electron microprobe analyzer for semiquantitative elemental analysis in energy-dispersive spectroscopy (EDS) mode, combing with the inductively coupled plasma-atomic emission spectrometer (ICP-AES, Teledyne Leeman Laboratories Prodigy 7).

Five spots in different areas were measured on one crystal using EDS and the ICP-AES measurement was performed on two pieces of single crystal. The crystal structure and chemical composition of $CsMn_6Bi_5$ single crystal have also been determined.

**Physical Property Measurements.** The magnetic susceptibility, resistivity, and heat capacity measurements were carried out using a physical property measurement system (PPMS, Quantum Design). Magnetic susceptibility was measured under a small magnetic field (0.1 T) parallel (H // rod) and perpendicular (H ⊥ rod) to the rod ([010] direction) using the zero-field-cooling (ZFC) and field-cooling (FC) protocols as reported in $RbMn_6Bi_5$ [16]. For larger fields at 1 T, 2 T, 3 T, 5 T, and 7 T, only magnetic susceptibility in FC protocol was measured. Magnetization hysteresis loops at 5 K, 10 K, 40 K, 50 K, 100 K, and 300 K were measured under the magnetic field up to 7 T parallel and perpendicular to the rod ([010] direction). The resistivity was measured using the standard four-probe configuration with the applied current (about 2 mA) parallel (I // rod) or perpendicular (I ⊥ rod) to the rod ([010] direction). Heat capacity measurement was carried out below 200 K. To protect the samples from air and moisture, thin film of N-type grease was spread to cover the sample for heat capacity measurement.

**First-Principles Calculations.** The first-principles calculations were carried out with the projector augmented wave (PAW) method as implemented in the Vienna *ab initio* simulation Package (VASP) [21, 22]. The generalized gradient approximation (GGA) [23] of the Perdew-Burke-Ernzerhof (PBE) [24] type was adopted for the exchange-correlation function. The cutoff energy of the plane-wave basis was 520 eV and the energy convergence standard was set to $10^{-6}$ eV. The $4 \times 4 \times 4$ and $2 \times 10 \times 4$ Monkhorst-Pack K-point meshes were employed for the Brillouin zone (BZ) sampling of the unit cell and the conventional cell, respectively.

3 Result and Discussion

**Crystal Structure.**

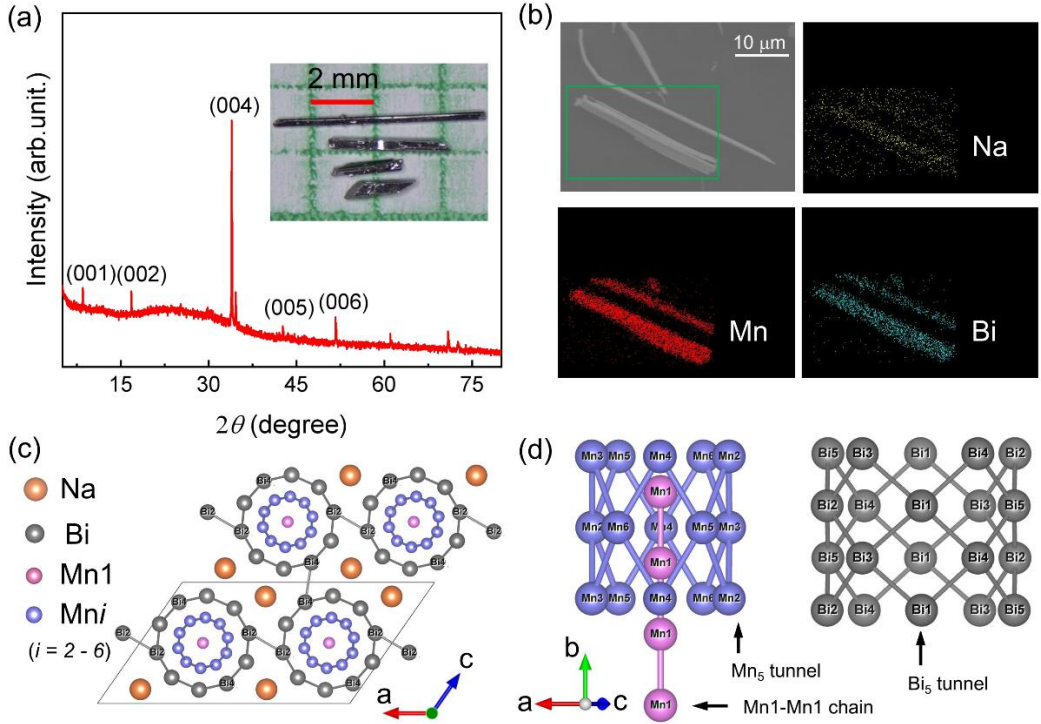

**Fig. 1** (a) X-ray diffraction pattern of an as-grown NaMn$_6$Bi$_5$ single crystal, showing (00$l$) reflections. The inset is the optical photograph of NaMn$_6$Bi$_5$ single crystals. (b) SEM image of NaMn$_6$Bi$_5$ and elemental mapping. (c) Crystal structure of NaMn$_6$Bi$_5$ viewed along $b$ axis with inter-column bonds Bi2-Bi2 and Bi4-Bi4 labeled. (d) Crystal structure of Mn$_5$ tunnel with one-dimensional Mn1-Mn1 chain at the center and Bi$_5$ tunnel viewed perpendicular to $ab$ plane.

As shown in the inset of Fig. 1a, the as-grown single crystals of NaMn$_6$Bi$_5$ are needle-like rods with shiny metal luster, indicating a clear quasi-one-dimensional feature. The X-ray diffraction pattern of an as-grown crystal is plotted in Fig. 1a with a strong preferential orientation of (00$l$) ($l$ = integer) reflections. Based on the position (~ 8.2°) of (001) diffraction peak, the distance between corresponding structural units was determined to be 12.6 Å, which is close to the lattice parameter $c$ as reported in KMn$_6$Bi$_5$ and RbMn$_6$Bi$_5$ [16, 17]. Elemental mapping of NaMn$_6$Bi$_5$ (Fig. 1b) confirms the homogeneous distribution of Na, Mn, and Bi. According to EDS (Fig. S1), the molar ratio is close to Na : Mn : Bi = 1 : 6.37(2) : 5.22(3), whereas ICP-AES indicates an molar ration of Na : Mn : Bi = 1.008(2) : 6.000(1) : 4.967(2). Considering its quasi-one-dimensional feature, we tried to mechanically exfoliate the crystals using the Scotch-tape method and obtained NaMn$_6$Bi$_5$ wires with diameters down to 528 nm (Fig. S1), showing more possibilities in applications.

The determined crystal structure based on the SCXRD data is shown in Fig. 1c and 1d. NaMn$_6$Bi$_5$ crystallizes in a monoclinic space group (Table S1) $C2/m$ (No. 11) with $a$ = 22.4357(25) Å, $b$ = 4.5952(5) Å, $c$ = 12.6577(13) Å, $\alpha = \gamma = 90°$, and $\beta = 123.109(3)°$. As shown in Fig. 1c, NaMn$_6$Bi$_5$ shares a same quasi-one-dimensional structure motif with that of KMn$_6$Bi$_5$ [17] and RbMn$_6$Bi$_5$ [17], featuring [Mn$_6$Bi$_5$]$^{-1}$ columns extending along the [010] direction, which is surrounded by the counter cation Na$^+$ acting as separators. The [Mn$_6$Bi$_5$]$^{-1}$ column is consisted of the Mn$_5$ tunnel built by Mn$i$ ($i$ = 2-6) with an inner one-dimensional Mn1-Mn1 atomic chain and the outmost Bi$_5$ tunnel.

**Table I**. Lattice parameters and typical bond lengths of AMn$_6$Bi$_5$ (A = Na, K, Rb, and Cs).

| A | Na | K [17] | Rb [16] | Cs |
|---|---|---|---|---|
| Radius (pm) | 102 | 138 | 152 | 167 |
| $a$ (Å) | 22.4357(25) | 22.994(2) | 23.286(5) | 23.6338(14) |
| $b$ (Å) | 4.5952(5) | 4.6128(3) | 4.6215(9) | 4.6189(3) |
| $c$ (Å) | 12.6577(13) | 13.3830(13) | 13.631(3) | 13.8948(8) |
| $\alpha, \gamma$ (°) | 90 | 90 | 90 | 90 |
| $\beta$ (°) | 123.1092(34) | 124.578(6) | 125.00(3) | 125.447(2) |
| Bi1-Bi3 (Å) | 3.5887 | 3.52 | 3.4931 | 3.4714 |
| Bi1-Bi4 (Å) | 3.6277 | 3.6115 | 3.6113 | 3.6015 |
| Bi2-Bi2 (Å) | 3.6358 | 3.5687 | 3.6061 | 3.6812 |
| Bi2-Bi4 (Å) | 3.5592 | 3.6517 | 3.7030 | 3.7698 |

The lattice parameters and typical bond lengths of AMn$_6$Bi$_5$ (A = Na, K, Rb, and Cs) are summarized in Table I. With the increasing cation radius from Na$^+$ (~102 pm) to K$^+$ (~138 pm), then to Rb$^+$ (~152 pm), and finally to Cs$^+$ (~167 pm), the lattice parameters $a$, $b$, $c$, and $\beta$ all show a monotonic increase. Compared with RbMn$_6$Bi$_5$, NaMn$_6$Bi$_5$ shows the shrinkage of $a$ and $c$ as large as 3.65% and 7.14%, respectively, whereas that of $b$ is only 0.57%. Such difference can be interpreted by the strong bonds of Mn-Mn and Bi-Bi along $b$ axis and the much weaker inter-column Na-Bi bonds. The decrease of $\beta$ suggests potential glides of [Mn$_6$Bi$_5$]$^{-1}$ columns along $a$ or $c$ axis, which means AMn$_6$Bi$_5$ (A = Na, K, Rb, and Cs) could be tuned under stress or pressure. Among all the bonds, the typical intra-column bond length monotonically decreases with increasing cation radius, from 3.5887 Å (Na) to 3.4931 Å (Cs) for Bi1-Bi3 and from 3.6277 Å (Na) to 3.6113 Å (Cs) for Bi1-Bi4, respectively. By contrast, the inter-column bond (Bi4-Bi4) monotonically increases from 3.5592 Å (Na) to 3.7698 Å (Cs) as cation radius increases, whereas another inter-column bond (Bi2-Bi2) shows an anomaly enhancement for NaMn$_6$Bi$_5$, with the relation 3.5687 Å (K) < 3.6061 Å (Rb) < 3.6358 Å (Na) < 3.6812 Å (Cs). The anomaly should be related with the unusual properties of NaMn$_6$Bi$_5$ (see below).

**Magnetic Property.**

Fig. 2a and 2b show the ZFC and FC magnetic susceptibility of NaMn$_6$Bi$_5$ single crystals under 0.1 T with $H$ // rod and $H \perp$ rod ([010] direction), respectively. The insets of Fig. 2a and 2b show the d$\chi$T/dT curves, donating the magnetic transition temperatures. For $H$ // rod, the ZFC and FC curves coincide with each other in the entire measured temperature range (2 K - 120 K), showing two AFM transitions at 45.3 K and 50.9 K. For $H \perp$ rod, a small deviation is observed in the AFM transitions. The bifurcation of ZFC and FC curves can be attributed to the magnetic anisotropy [25]. Compared with transition temperatures of KMn$_6$Bi$_5$ (~75 K) [17] and RbMn$_6$Bi$_5$ (~80 K) [16], the AFM transition is largely suppressed in NaMn$_6$Bi$_5$. Moreover, the AFM transition splits. As predicted in RbMn$_6$Bi$_5$, the moments of Mn1 at the center couple with the moments of Mn$i$ ($i$ = 2-6) at the pentagon, inducing a helical magnetic structure with one magnetic transition temperature [16]. Once these two kinds of moments are decoupled, multiple AFM transitions are expected. The observed lower AFM transition temperatures and split AFM transitions may be the hint of such decoupling, whereas the underline mechanism still remains uncovered. More importantly, the lower AFM transition temperature makes NaMn$_6$Bi$_5$ a more promising candidate to explore Mn-based superconductor by chemical doping or applying physical pressure.

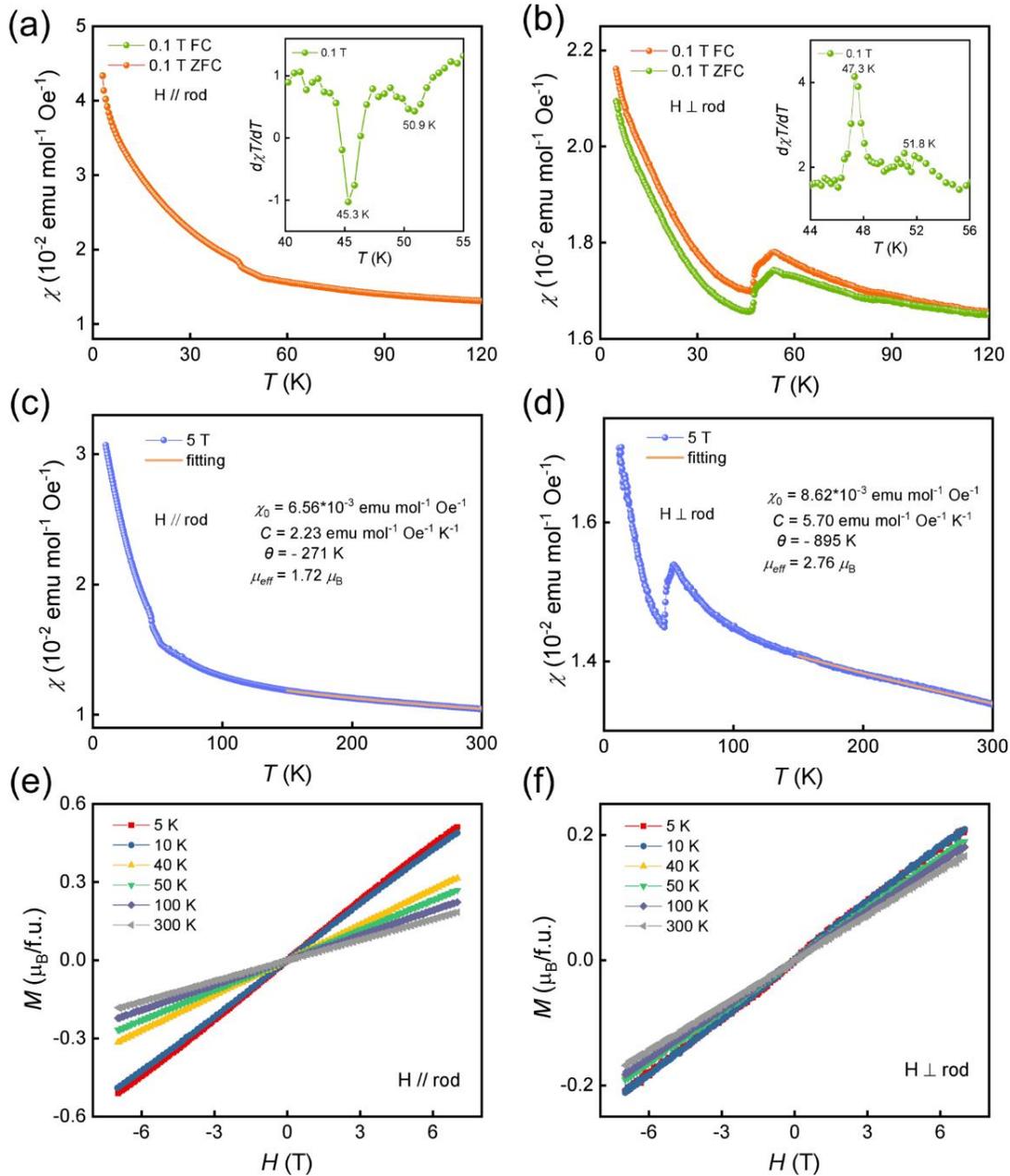

**Fig. 2** ZFC and FC magnetic susceptibility of NaMn$_6$Bi$_5$ single crystals under 0.1 T for (a) *H* // rod and (b) *H* ⊥ rod ([010] direction). The insets are the corresponding d$\chi$T/dT curves. Temperature dependent magnetic susceptibility of NaMn$_6$Bi$_5$ single crystals under 5 T for (c) *H* // rod and (d) *H* ⊥ rod and corresponding Curie-Weiss fittings from 150 K to 300 K. Field dependent magnetizations for (e) *H* // rod and (f) *H* ⊥ rod at different temperatures.

By fitting the magnetic susceptibility under 5 T at high temperature range (150 K-300 K) using Curie-Weiss law, $\chi = \chi_0 + \frac{C}{T-\theta}$, where $\chi_0$ is the temperature-independent contribution including the diamagnetic contribution of the orbital magnetic moment and the Pauli paramagnetic contribution of conduction electron, $C$ is the Curie constant, and $\theta$ is the Curie temperature [26]. The fitted values are $\chi_0 = 6.56 \times 10^{-3}$ emu mol$^{-1}$ Oe$^{-1}$, $\theta = -271$ K, and $C = 2.23$ emu mol$^{-1}$ Oe$^{-1}$ K$^{-1}$ for $H$ // rod, whereas $\chi_0 = 8.62 \times 10^{-3}$ emu mol$^{-1}$ Oe$^{-1}$, $\theta = -895$ K, and $C = 5.70$ emu mol$^{-1}$ Oe$^{-1}$ K$^{-1}$ for $H \perp$ rod. Negative $\theta$ indicates that the interactions in both directions are of AFM character. The frustration factors $|\theta/T_N|$ [27] are calculated to be ~6 for $H$ // rod and ~18 for $H \perp$ rod, indicating an anisotropic frustration. This may be attributed to the quite large magnetic anisotropy resulted from the geometry of the quasi-one-dimensional motif.

The effective moment can be derived following the equation $\mu_{eff} = \sqrt{\frac{8C}{n}}$, where $n$ is the number of magnetic atoms. For NaMn$_6$Bi$_5$, the effective moment is $\mu_{eff}$ ~ 1.72 $\mu_B$/Mn for $H$ // rod and $\mu_{eff}$ ~ 2.76 $\mu_B$/Mn for $H \perp$ rod, respectively. As discussed in RbMn$_6$Bi$_5$ [16], these values fall in the range of the spin-only magnetic moment of low-spin $t_{2g}^{4.67}$ (1.38 $\mu_B$) and high-spin $t_{2g}^{3}e_g^{1.67}$ (5.48 $\mu_B$) of octahedrally coordinated d$^{4.67}$, corresponding to the Zintli phase [28] with an average oxidation state of Mn$^{2.33+}$. The magnetic anisotropy in NaMn$_6$Bi$_5$ is larger than that of KMn$_6$Bi$_5$, which maybe be attributed to the effect of cation radius.

The magnetization of NaMn$_6$Bi$_5$ under magnetic field parallel and perpendicular to the rod ([010] direction) are shown in Fig. 2e and 2f, respectively. All magnetization curves increase linearly with magnetic field ranging from 5 K to 300 K and remain unsaturated up to 7 T with no hysteresis loop in both directions. The absence of hysteresis in M-H curves confirms the dominant AFM interaction at low temperatures, same as in KMn$_6$Bi$_5$ [17] and RbMn$_6$Bi$_5$ [16]. By increasing the magnetic field from 1 T to 7 T for $H$ // rod, the two AFM transition temperatures hardly change (Fig. S2), showing the Mn moments in NaMn$_6$Bi$_5$ having large saturation magnetic field.

**Resistivity.**

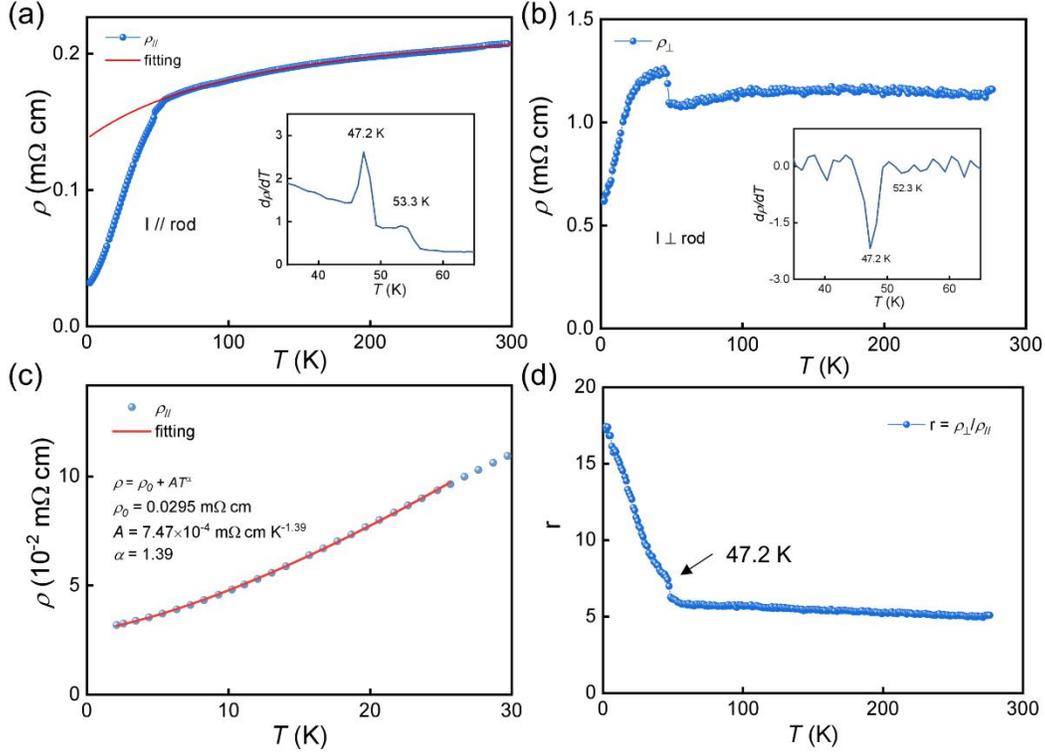

**Fig. 3** Temperature dependent resistivity $\rho(T)$ for NaMn$_6$Bi$_5$ single crystal measured with (a) $I \parallel$ rod and (b) $I \perp$ rod ([010] direction). The insets are the first derivative in the temperature range 35 K - 65 K. (c) The enlarged low-temperature resistivity with $I \parallel$ rod. (d) Temperature dependent anisotropic resistivity ratio for NaMn$_6$Bi$_5$ single crystal.

With $I \parallel$ rod, the resistivity of NaMn$_6$Bi$_5$ decreases with decreasing temperature, showing a metallic-like behavior with an anomaly around 50 K (Fig. 3a). From the first derivative curve (d$\rho$/dT), two transition temperatures at 47.2 K and 53.3 K can be extracted and coincide with AFM transition temperatures observed in the magnetic susceptibility, which are both direction-independent (Fig. 3b) and sample-independent (Fig. S3). The room-temperature resistivity for NaMn$_6$Bi$_5$ is estimated to be 0.207 m$\Omega$ cm in the [010] direction, smaller than the maximum value (1 m$\Omega$ cm) determined by the MIR limit [29]. The residual resistivity ratio RRR = $\frac{\rho(300\ \text{K})}{\rho(2\ \text{K})}$ = 6.5 indicates that the crystalline quality of NaMn$_6$Bi$_5$ is good. Above 55 K, the resistivity can be fitted (red line in Fig. 3a) using the formula $\frac{1}{\rho(T)} = \frac{1}{\rho_{\text{sat}}} + \frac{1}{\rho_{\text{ideal}}}$, where $\rho_{\text{sat}}$ represents the saturation resistivity, $\rho_{ideal}$ is the ideal resistivity that satisfies the Boltzmann equation. $\rho_{ideal}$ is proportional to the temperature in the high temperature region and can be described by $\rho_{ideal} = \rho_r + aT$ [29]. The fitting yields $\rho_{sat} = 0.24$ mΩ cm, $\rho_r = 0.33$ mΩ cm, and $a = 4.4 \times 10^{-3}$ mΩ cm K$^{-1}$. As shown in Fig. 4c, the resistivity can be well fitted using $\rho = \rho_0 + AT^{\alpha}$ below 25 K, with fitted parameters $\rho_0 = 0.0295$ mΩ cm, $A = 7.47 \times 10^{-4}$ mΩ cm K$^{-1.39}$, and $\alpha = 1.39$. The value of power $\alpha$ deviates largely from 2, suggesting a non-Fermi liquid behavior of NaMn$_6$Bi$_5$ at low temperature, which is quite different from the Fermi liquid behavior reported in both KMn$_6$Bi$_5$ ($\alpha \sim 2$) [17] and RbMn$_6$Bi$_5$ ($\alpha = 1.9$) [16].

In comparison, temperature dependent resistivity with $I \perp$ rod shows an increase below the AFM transition temperature (Fig. 3b). After reaching the maximum, the resistivity begins to decrease again with the decreasing temperature. Similar to that in KMn$_6$Bi$_5$ and RbMn$_6$Bi$_5$, the jump is attributed to a dimensional crossover of electrons in low-dimensional materials [16, 17, 30]. The anisotropic resistivity ratio, donated as r = $\rho_\perp/\rho_{//}$, increases with decreasing temperature and sharply increases below the AFM transition temperature (Fig. 3d). The value of anisotropic resistivity ratio is about 17 at 2 K, comparable to that of KMn$_6$Bi$_5$ (~20) [17] but much smaller than that of RbMn$_6$Bi$_5$ (~240) [16].

**Specific Heat Capacity.**

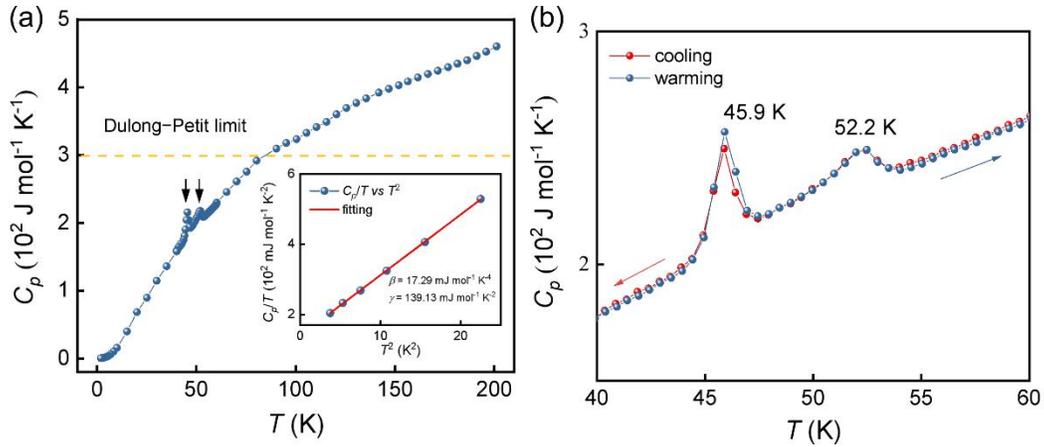

**Fig. 4** (a) Temperature dependent specific heat capacity for NaMn$_6$Bi$_5$ single crystal. The inset shows the $C_p/T$ versus $T^2$, the red solid line is the linear fitting by using Debye model. (b) The enlarged specific heat capacity around 50 K for cooling down and warming up.

Temperature dependent specific heat capacity for NaMn$_6$Bi$_5$ single crystal measuring from 2 K to 200 K is shown in Fig. 4a, which exhibits two peaks at 45.9 K and 52.2 K. These temperatures are close to the transition temperatures of magnetic susceptibility and resistivity, which are sample-independent and reversible for cooling down and warming up (Fig. 4b). The specific heat capacity exceeds the Dulong–Petit limit (3NR ~ 300 J mol$^{-1}$ K$^{-1}$, yellow dash line) at high temperature, which is also attributed to more prominent phonon contribution at higher temperature of N-type grease used for protecting the sample [16, 17]. For NaMn$_6$Bi$_5$, both jumps $\Delta C$ ~ 28 J mol$^{-1}$ K$^{-1}$ and 11 J mol$^{-1}$ K$^{-1}$ are much smaller than that of RbMn$_6$Bi$_5$ ($\Delta C$ ~ 175 J mol$^{-1}$ K$^{-1}$) and KMn$_6$Bi$_5$ ($\Delta C$ ~ 180 J mol$^{-1}$ K$^{-1}$). We then fit the low-temperature (2 K - 5 K) specific heat capacity with Debye model $C = \gamma T + \beta T^3$, where $\gamma T$ represents the contribution from the electron, $\beta T^3$ represents the contribution of the lattice. The Sommerfeld coefficient $\gamma$ proportional to the DOS at the Fermi level is determined to be 139.13 mJ K$^{-2}$ per formula [23.2 mJ K$^{-2}$ (mol-Mn)$^{-1}$], much larger (3~5 times) than that of RbMn$_6$Bi$_5$ (28.2 mJ K$^{-2}$ per formula) [16] and KMn$_6$Bi$_5$ (39.0 mJ K$^{-2}$ per formula) [17]. This indicates the decreasing trend of DOS at the Fermi level by increasing the cation radius from Na to Rb, which is also held for other quasi-one-dimensional materials like A$_2$Cr$_3$As$_3$ (A = K, Rb, and Cs) [12-14] with the underlying mechanism unknown. The Debye temperature $\Theta_D$ can be calculated by $(12\pi^4 Rn/5\beta)^{1/3}$, where $R$ represents the molar gas constant and $n$ is the number of atoms in the chemical formula. For NaMn$_6$Bi$_5$, the Debye temperature $\Theta_D$ is 110.8 K, quiet close to that of RbMn$_6$Bi$_5$ (107.8 K) [16] and KMn$_6$Bi$_5$ (114.6 K) [17], which can be attributed to the quite similar crystal structure of these compounds.

**Density of States.**

We then tried to understand the effect of structure anomaly in NaMn$_6$Bi$_5$ and the origin of its different magnetic transitions, non-Fermi liquid behavior, and much enhanced Sommerfeld coefficient by calculating DOS near the Fermi level for AMn$_6$Bi$_5$ (A = Na, K, Rb, and Cs) using first-principles calculations. As shown in Fig. 5, all AMn$_6$Bi$_5$ (A = Na, K, Rb, and Cs) compounds have non-zero DOS at Fermi level ($E_F$), corresponding well with the observed metallic-like behavior in resistivity. Compared with other $AMn_6Bi_5$ (A = K, Rb, and Cs) compounds, $NaMn_6Bi_5$ shows a quite different DOS near the Fermi level ($E_F$), which may correspond to those different characteristics. Specially, a peak and a larger dip (red arrows) emerge around $E_F$ in $NaMn_6Bi_5$, which should be responsible for the much enhanced Sommerfeld coefficient.

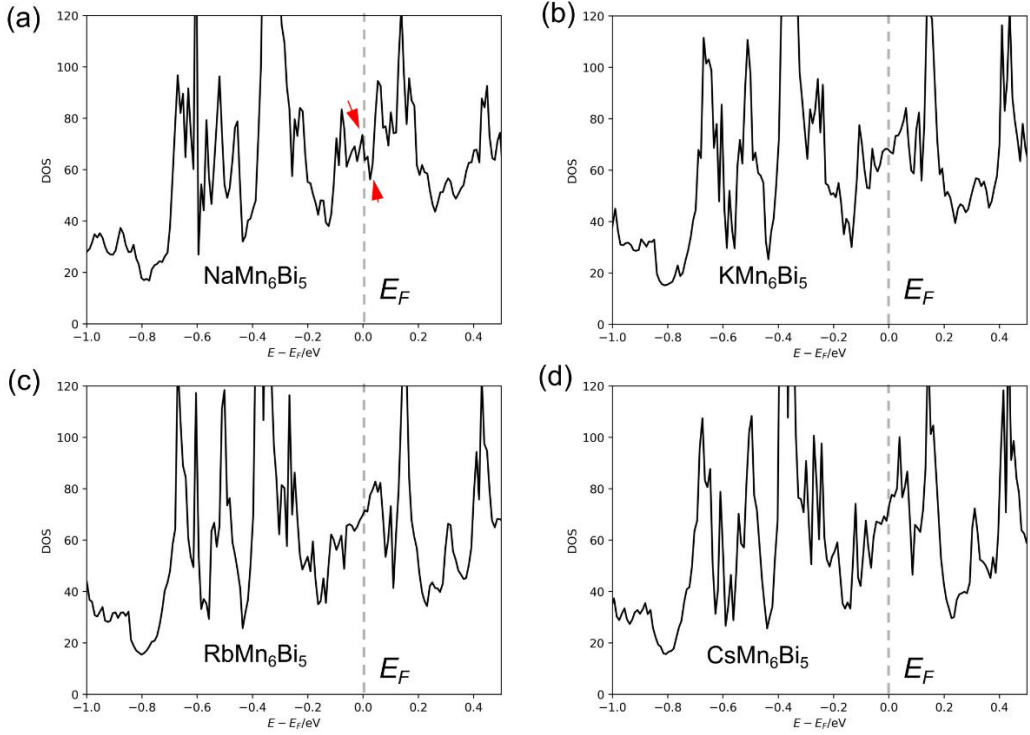

**Fig. 5** DOS near the Fermi level ($E_F$, dash lines) from -1 eV to 0.5 eV for (a) $NaMn_6Bi_5$, (b) $KMn_6Bi_5$, (c) $RbMn_6Bi_5$, and (d) $CsMn_6Bi_5$. Red arrows donate the difference of DOS for $NaMn_6Bi_5$ around $E_F$ compared with other $AMn_6Bi_5$ (A = K, Rb, and Cs) compounds.

## 4 Conclusion

In conclusion, a new quasi-one-dimensional superconductor parent compound $NaMn_6Bi_5$ has been discovered. Compared with other $AMn_6Bi_5$ (A = K, Rb, and Cs), $NaMn_6Bi_5$ has a larger intra-column Bi-Bi bond length. This structure anomaly may result in the two decoupled AFM transitions at 47.3 K and 51.8 K observed in magnetic susceptibility, resistivity, and heat capacity measurements. Anisotropic resistivity, non-Fermi liquid behavior, and much enhanced Sommerfeld coefficient are observed, which may be attributed to the different DOS near the Fermi level compared with other $AMn_6Bi_5$ (A = K, Rb, and Cs) compounds. The decoupled AFM transitions and lower AFM transition temperatures make $NaMn_6Bi_5$ a more promising platform for exploring the exchange-coupling interaction and superconductivity. Chemical doping and physical pressure should be the effective means to induce superconductivity within such quasi-one-dimensional materials with possible helical magnetic order.

**Supporting information**

Table S1 compares the crystal structure of $NaMn_6Bi_5$ and $CsMn_6Bi_5$. Figure S1 displays the quasi-one-dimensional feature of $NaMn_6Bi_5$. Figure S2 shows the magnetic susceptibility of $NaMn_6Bi_5$ single crystal under various magnetic field. Figure S3-S4 are temperature dependent resistivity and specific heat capacity of other $NaMn_6Bi_5$ single crystals, respectively.

**Author information**

**Notes**

The authors declare no competing financial interest.

**Acknowledgements**

This work was partially supported by the National Key Research and Development Program of China (Grant Nos. 2017YFA0302902 and 2018YFE0202600), the National Natural Science Foundation of China (Grant No. 51832010), , and the Key Research Program of Frontier Sciences, Chinese Academy of Sciences (Grant No. QYZDJ-SSW-SLH013).

## Supporting information

**Table S1.** Crystal structure of NaMn$_6$Bi$_5$ and CsMn$_6$Bi$_5$ obtained by structural refinement from single crystal X-ray diffraction data.

| Empirical formula | NaMn$_6$Bi$_5$ | CsMn$_6$Bi$_5$ |
|---|---|---|
| Formula weight | 1397.52 g/mol | 1507.43 g/mol |
| Space group / Z | $C2/m$ (No. 12) /4 | $C2/m$ (No. 12) /4 |
| Unit cell dimensions | $a$ = 22.581(6) Å | $a$ = 23.6338(14) Å |
|  | $b$ = 4.6207(13) Å | $b$ = 4.6189(3) Å |
|  | $c$ = 12.754 (4) Å | $c$ = 13.8948 (8) Å |
|  | $\alpha = \gamma = 90°$ | $\alpha = \gamma = 90°$ |
|  | $\beta$ = 123.196(8)° | $\beta$ = 125.447(2)° |
| Volume / $d_{cal}$ | 1113.6(6) Å$^3$ / 8.961 g/cm$^3$ | 1235.6(5) Å$^3$ / 8.711 g/cm$^3$ |
| Reflections collected/R(int) | 5362 / 0.1418 | 6427 / 0.079 |
| Data / restraints / parameters | 1499 / 0 / 73 | 1278 / 0 / 73 |
| Final R indices [I > 2sigma(I)] | $R_1$ = 0.0615, $wR_2$ = 0.1607 | $R_1$ = 0.0512, $wR_2$ = 0.0565 |
| R indices (all data) | $R_1$ = 0.0872, $wR_2$ = 0.1951 | $R_1$ = 0.0595, $wR_2$ = 0.0603 |
| Largest diff. peak and hole | 2.010 and -2.491 e.Å$^{-3}$ | 1.010 and -2.132 e.Å$^{-3}$ |

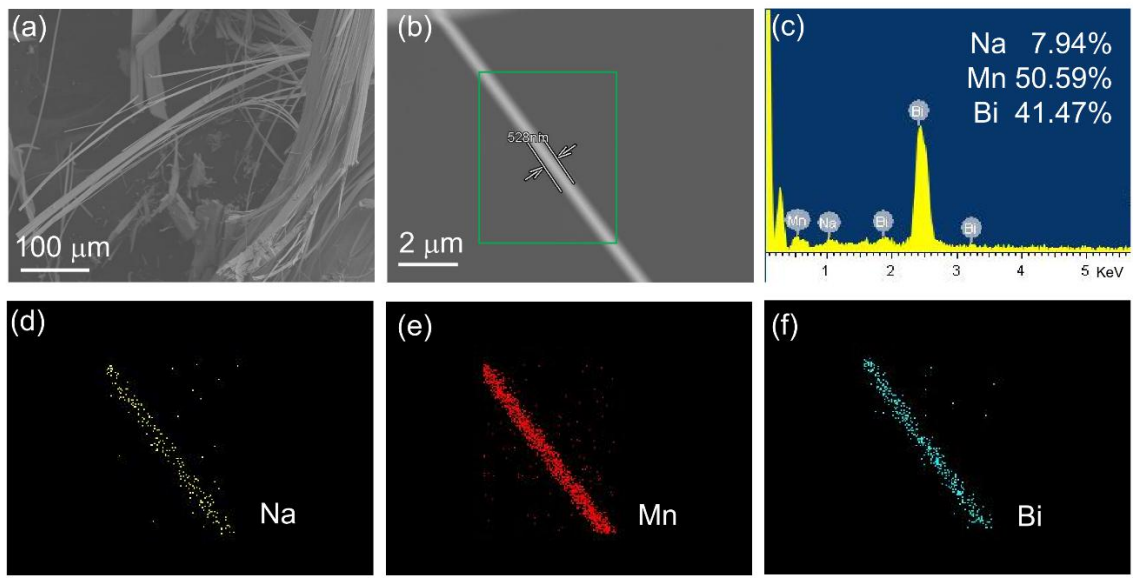

**Fig. S1** Quasi-one-dimensional feature of NaMn$_6$Bi$_5$. SEM image of NaMn$_6$Bi$_5$ at a scale of (a) 100 μm and (b) 2 μm. Elemental analysis (c) and elemental mapping for (d) Na, (e) Mn, and (f) Bi by EDX on the wire.

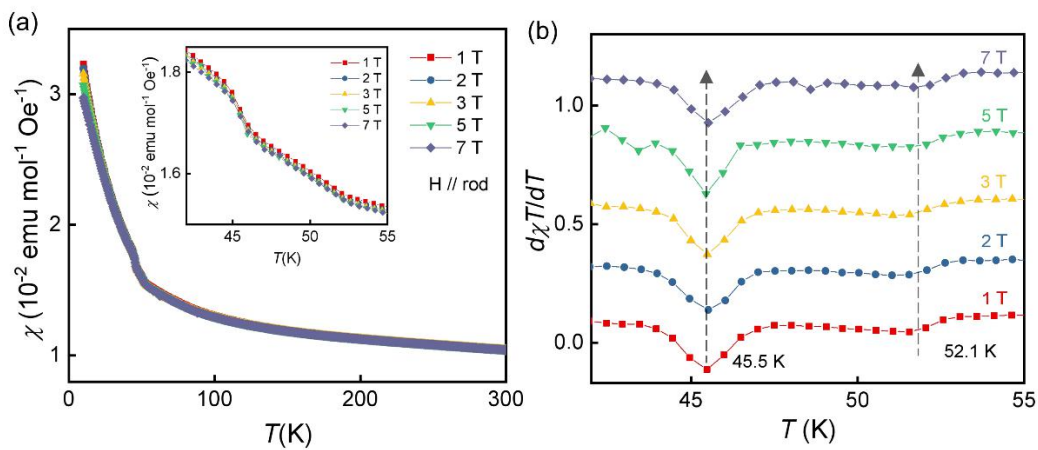

**Fig. S2** (a) Temperature dependent magnetic susceptibility of NaMn$_6$Bi$_5$ single crystal under various magnetic field for $H \parallel$ rod. The inset is the zoomed susceptibility around 50 K. (b) The $d\chi T/dT$ curves in the temperature range 42 K - 55 K, showing almost unchanged magnetic transition temperatures with increasing magnetic field.

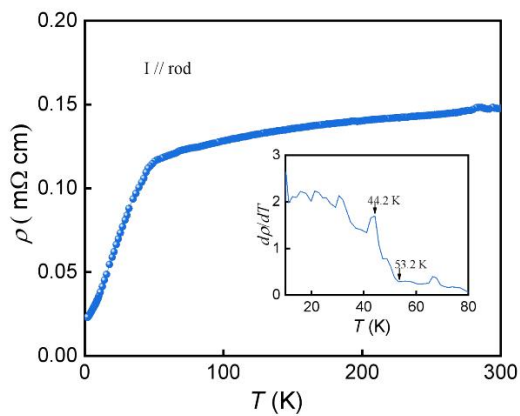

**Fig. S3** Temperature dependent resistivity of another NaMn$_6$Bi$_5$ single crystal. The inset is the first derivative curve in the temperature range 10 K - 80 K.

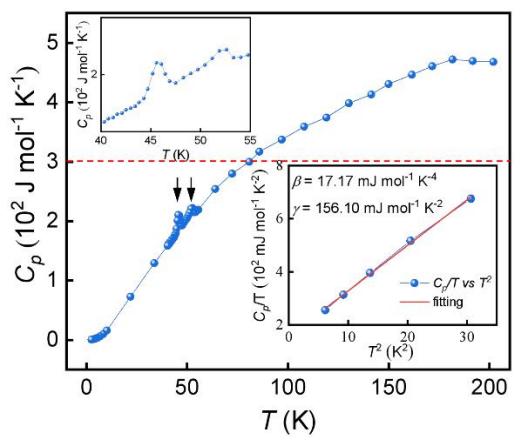

**Fig. S4** Temperature dependent specific heat capacity of another NaMn$_6$Bi$_5$ single crystal and the corresponding linear fitting at low temperature range by using Debye model.